\documentclass[11pt,a4paper]{article}

\usepackage{jcappub}
\usepackage[margin=0.95in]{geometry}

\usepackage[utf8]{inputenc}
\usepackage{graphicx}
\usepackage{subcaption}
\usepackage[export]{adjustbox}
\usepackage{amsmath}
\usepackage{hyperref}

\newcommand{\nn}{\nonumber}

\def\MR{{{\cal M}}_{\rm I}}
\def\ML{{\cal M}_{{\rm II}}}
\def\dA{\dot{A}}
\def\dB{\dot{B}}

\newcommand{\be}{\begin{equation}\begin{aligned}}
\newcommand{\ee}{\end{aligned}\end{equation}}
\newcommand{\bbe}{\begin{align}}
\newcommand{\eee}{\end{align}}
\newcommand{\bea}{\begin{eqnarray}}
\newcommand{\eea}{\end{eqnarray}}
\def\beq{\begin{equation}}
\def\eeq{\end{equation}}

\def\lsim{\mathrel{\rlap{\lower4pt\hbox{\hskip0.5pt$\sim$}}
		\raise1pt\hbox{$<$}}}     
\def\gsim{\mathrel{\rlap{\lower4pt\hbox{\hskip0.5pt$\sim$}}
		\raise1pt\hbox{$>$}}}     
	
	\def\d{{\rm d}}

\title{  Topological Early Universe Cosmology }

\author[a]{A. Kehagias,}
\author[b]{A. Riotto}

\affiliation[a]{Physics Division, National Technical University of Athens, Zografou, Athens, 15780, Greece} 
\affiliation[b]{Department of Theoretical Physics and Center for Astroparticle Physics (CAP) \\
			24 quai E. Ansermet, CH-1211 Geneva 4, Switzerland}

\abstract{
The early
 history of the universe might  be described by a 
topological phase followed by a standard second phase of Einstein gravity. 
To study  this scenario  in its full generality, we  consider  a   four-manifold of Euclidean signature 
in the topological phase, which shares a common boundary with a corresponding manifold of Lorentzian signature in the  Einstein phase. We find that the boundary should have vanishing extrinsic curvature, whereas the manifold in the topological  phase should have zero Euler number.  
In addition, we show that the second phase must be characterized by an initial  vanishing Weyl tensor and that   the  standard cosmological  flatness problem 
is not automatically solved unless a  conformal invariant boundary term is added.
 We also characterize the  scalar perturbations  in the standard Einstein phase.  We  show that they must contain an initial  non-vanishing shear component inherited from the topological phase and 
we estimate the non-Gaussian parameters.
Finally, we argue that the   topological early universe cosmology    
shares common features of
  previous ideas, such as the so-called  Weyl curvature hypothesis,  the universe's creation out of nothing  and the no-boundary proposal. 
}

\emailAdd{kehagias@central.ntua.gr}
\emailAdd{antonio.riotto@unige.ch}
\begin{document}

\maketitle
\flushbottom

\section{Introduction and Conclusions}
It has been recently proposed that the early phase of the universe may not be governed by Einstein gravity,  but rather described by a topological phase \cite{Vafa}. According to this idea, partially motivated by string theory and duality symmetries, the matter content of the universe could be quite different from the one we explore today. For example, in  string gas cosmology \cite{BrV} the matter of the universe is  described by momentum modes which are  very heavy  when the radius of the universe is at the string scale. On the other hand, winding modes (normally superheavy at large scales) are light at that scale, and they  replace the momentum modes for the matter content  driving the dynamics. The new idea proposed in Ref. \cite{Vafa} is that, viewed from our current perspective, the early phase of the universe is described by a topological phase during which the  gravitational degrees of freedom are absent as there are no metric fluctuations. In this phase, Einstein gravity is replaced by Witten's topological gravity \cite{Witten} on a 4D Riemannian manifold of Euclidean signature. This theory is a kind  of  conformal gravity 
with a local fermionic symmetry of  BRST type. It does not have local degrees of freedom since all degrees of freedom in this topological phase  (phase I)  can be gauged away due to the BRST invariance. Local physics emerges due to  anomalies. The local diffeomorphism invariance is broken down to Poincar\'e symmetry in the second, non-topological Einstein phase (phase II).
 The latter is 
described by a 4D pseudo-Riemannian manifold of Lorentzian signature with dynamics determined by the Einstein equations. The two phases are sewed together at a common boundary hypersurface $B$, much like the continuation from 
the Euclidean to Lorentzian regime in the no-boundary proposal \cite{Hartle-Hawking}  and  the creation of the universe from nothing \cite{VilenkinI,VilenkinII}. From this point of view, the topological phase  provides initial conditions on the Cauchy hypersurface at some time $t=0$ for the evolution of the universe under Einstein equations in phase II. 

%
In this paper we elaborate on the topological early universe cosmology, by characterizing the physical conditions emerging in the present setup at the boundary. Our results can be summarized as follows.

\begin{enumerate}

\item Phase I may be described by a Riemannian manifold of Euclidean signature $\MR$ and self-dual Weyl tensor,  and phase II by  a pseudo-Riemannian manifold $\ML$ of Lorentzian signature. The two phases have a common boundary $B$. The manifold $B$ is a codimension one totally geodesic  submanifold (vanishing extrinsic  curvature). The  Euler number of $\MR$ vanishes, while    
   the scalar curvature of  $B$ is non-negative. 
   
 \item  The standard cosmological horizon problem  can be solved, but the flatness problem requires non-trivial dynamics on the boundary hypersurface.  
   
   \item The initial value of Weyl tensor  in phase II vanishes.

   \item 
Scalar perturbations are created due to the breaking of conformal symmetry
by the trace anomaly and there are no tensor perturbations. However, the matching at the boundary requires a non-vanishing initial shear in phase II. The amount of non-Gaussianity of the scalar perturbations in the squeezed limit \cite{ngreview}  turns out to be parametrized by $f_{NL}={\cal O}(1)$ as far as the three-point correlator is concerned.

\item The  topological early universe cosmology   is related one way or the other with previous approaches and in particular with the ``Weyl curvature hypothesis" \cite{PenroseI,PenroseII,PenroseIII}, the universe's creation out of nothing \cite{VilenkinI,VilenkinII} and the no-boundary   proposal \cite{Hartle-Hawking}. 

\end{enumerate}

\noindent
The structure of the paper is the following: in section 2, we describe the two-phase model. In section 3 we discuss the cosmological problems in the present framework. In section 4 we consider cosmological perturbations
and in section 5 we compare the two-phase model with previous proposals.

\section{The two-phase gravity model}
We will assume as in Ref. \cite{Vafa}, that gravity  appears in two phases and is described  by the geometry of a   4D manifold ${\cal M}$.
 We will denote the manifold in phase I as ${\cal M}_{\rm I}$ and the manifold in phase II as ${\cal M}_{\rm II}$. Therefore, if $B$ is the common boundary of $\MR$ and $\ML$, we may write 
\begin{align}
{\cal M}&=\MR\cup \ML, \nonumber \\
\partial \MR&=\partial\ML=B. \label{M}
\end{align}
\begin{figure}[h!]
\includegraphics[angle=-90,width=13cm]{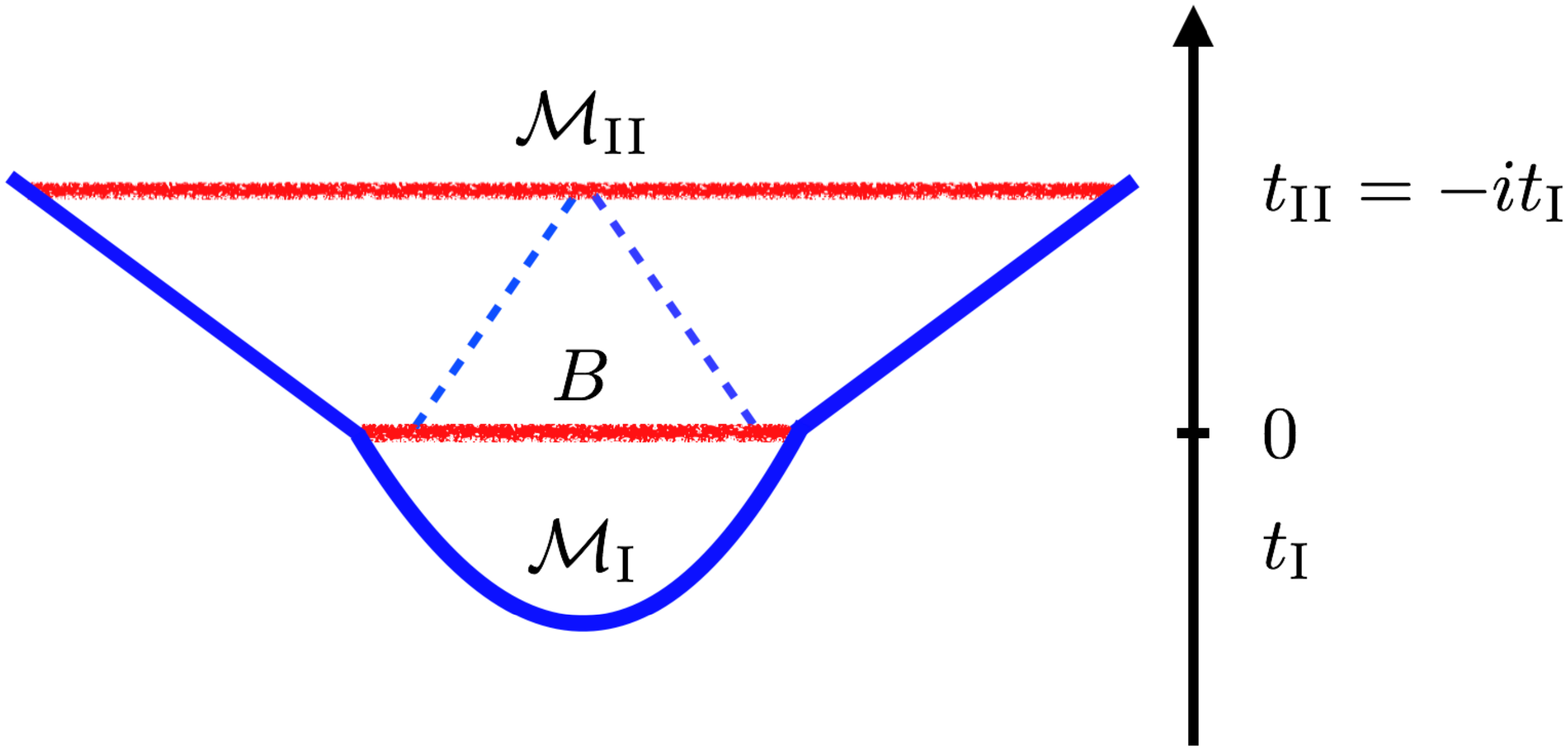}
\centering
\caption{{\small The two manifolds $\MR$ and $\ML$ are connected through their common boundary $B$. $\MR$ and $\ML$ are endowed with  Riemannian and Lorentzian metrics, respectively.  The time coordinate $t_{\rm II }$ in $\ML$ is the Wick rotation of $t_{\rm I}$ in $\MR$. The geometry is of the ``no-boundary'' type.}}
\end{figure}
In particular, the manifold $\MR$ has a  Riemannian metric of $(+,+,+,+)$ signature, whereas 
$\ML$ has a Lorentzian metric with signature $(-,+,+,+)$. The induced metric $\gamma_{ij}$ ($i,j=1,2,3)$ either from  the $\MR$ or the  $\ML$ side  agree on their common  boundary $B$
\begin{eqnarray}
 \gamma_{ij}\Big|_{\partial \MR}=\gamma_{ij}\Big|_{\partial \ML}.
 \end{eqnarray} 
The gravitational dynamics in $\ML$ is governed
by the Einstein equations
\begin{eqnarray}
G_{\mu\nu}=R_{\mu\nu}-\frac{1}{2}g_{\mu\nu} R=M_{\rm p}^{-2}T_{\mu\nu}, ~~~~\mu,\nu=0,1,2,3, \label{EE}
\end{eqnarray}
where $M_{\rm p}$ is the Planck mass and $T_{\mu\nu}$ is the energy-momentum tensor. 
On the other hand, the theory in $\MR$ is assumed to be  topological as advocated in  Ref. \cite{Vafa}, and  in particular, it is Witten's topological gravity.  The action of  the pure gravitational part of the latter is the conformal Weyl square  theory
\begin{eqnarray}
{\cal S}_{\rm top}= \int_{{\cal M}_{\rm I}} \d^4 x \sqrt{g}\,  \left(\frac{1}{2 g_{\tiny{W}}^2} W_{\mu\nu\rho\sigma}
W^{\mu\nu\rho\sigma}+\frac{1}{2 g_{\tiny{E}}^2} E_4\right), \label{actionI}
\end{eqnarray}
where  $W_{\mu\nu\rho\sigma}$ is the Weyl tensor,
\begin{eqnarray}
E_4=R_{\mu\nu\rho\sigma}R^{\mu\nu\rho\sigma}-4 R_{\mu\nu}R^{\mu\nu}+R^2
\end{eqnarray}
is the Euler density and  $g_W$ and $g_E$ are the corresponding couplings. Usally, the $E_4$ term is not written as it is a topological term, but we keep it as the theory is topological anyway. 
 The full action is given by
\begin{eqnarray}
{\cal S}_{I}= {\cal S}_{\rm top}+{\cal S}_{\rm KS},   \label{SI}
\end{eqnarray}
where 
\begin{eqnarray}
{\cal S}_{\rm KS}=
\int_{{\cal M}_{\rm I}} {\d^4} x \sqrt{-g}
\left\{ \tau \Big(c W_{\mu\nu\rho\sigma}W^{\mu\nu\rho\sigma}-a E_4\Big)-4 a \,\partial_\mu \tau\partial_\nu \tau \left(G^{\mu\nu}-g^{\mu\nu}\Box \tau+\frac{1}{2}\partial^\mu \tau\partial^\nu\tau\right)\right\}\nonumber\\
\end{eqnarray}
  is  
the trace-anomaly  action \cite{riegert0,K-S} and    the scalar $\tau$ is the dilaton.
The full action (\ref{SI})  contains second derivatives of the metric and therefore it is a higher-derivative theory. Such theories are  classically unstable and quantum mechanically have indefinite metric Hilbert space with ghosts. Indeed,
 the fourth order equations of motion of Weyl conformal  gravity give rise not only to the ordinary massless graviton but also  to helicity $(\pm 2, \pm 1,0)$ ghost-like states  \cite{Stelle}, forming all together  a helicity $\pm2$  dipole ghost \cite{FZ,FGvN,LvN,Riegert,AKKLR,FKL}.  In addition, there exists  an ordinary massless helicity $\pm 1$ vector in the spectrum. 
Therefore, despite of its improved UV properties, it is questionable if the theory (\ref{actionI})  can make sense at all as a consistent gravity theory since  a  way to eliminate the ghosts  from conformal gravity is lacking until to now. 
Supersymmetry cannot help here since it just adds  extra fermionic ghost degrees of freedom.  However, Witten implemented a 
BRST symmetry to the theory by adding new   bosonic and fermionic degrees of freedom which eliminates not only the ghosts but all degrees of freedom  altogether. In other words, in Witten's topological theory
there are no propagating degrees of freedom. 

Fields in $\MR$ are in $SO(4)=SU(2)\otimes SU(2)$ representations. 
In a two-component spinor notation, a field $\Psi_{A_1,\cdots,A_m,\dA_1,\cdots,\dA_n}$ has spin $(m/2,n/2)$, and an $n$-index tensor $\Psi_{\mu_1,\cdots,\mu_n}$ can be written as $\Psi_{A_1,\cdots,A_n,\dA_1,\cdots,\dA_n}$. The latter can be decomposed in irreducible representations by appropriate symmetrizations and antisymmetrization of its indices. 
The field content of Witten's topological gravity contains the metric (vierbien) $e_{\mu A\dA}$ and two additional bosonic fields $B_{A\dA}$ and $C_{A\dA}$ in addition to the fermions $\lambda_{A\dA}$, $\psi_{AB\dA\dB}$ and $\chi_{ABCD}$. The corresponding fermionic  BRST shifts are 
\begin{eqnarray}
\delta \lambda_{A\dA}&=&\epsilon C^\mu D_\mu \lambda_{A\dA}+\frac{i}{2}
\epsilon\Big(\Lambda_{AC} {B^{C}}_{\dA}+\Lambda_{\dA\dot{C}} {B^{\dot{C}}}_A\Big)
+\frac{1}{2}\epsilon k S B_{A\dA}, \nn \\
\delta \psi_{AB\dA\dB}&=&\frac{1}{2}\epsilon \Big(e_{\mu A\dA}D^\mu C_{B\dB}+
e_{\mu B\dA}D^\mu C_{A\dB}+e_{\mu A\dB}D^\mu C_{B\dA}+e_{\mu B\dB}D^\mu C_{A\dA}\Big) , \nn \\
\delta \chi_{ABCD}&=& i W_{ABCD}, \label{delta}
\end{eqnarray}
where $k$ is the conformal dimension of $\lambda$, $\Lambda_{AB}$ and  $S$ are appropriate  quadratic expressions of the bosonic fields \cite{Witten}, and $W_{ABCD}$ is the spin $(0,2)$ content of the Weyl tensor, i.e.,  its self dual part. Note that the Weyl tensor $W_{\mu\nu\rho\sigma}$ in $\MR$   can be decomposed into a self-dual and an anti-self dual part as 
\begin{eqnarray}
W_{\mu\nu\rho\sigma}&=&W^+_{\mu\nu\rho\sigma}+W^-_{\mu\nu\rho\sigma} ,
\nn\\
W^\pm_{\mu\nu\rho\sigma}&=&\frac{1}{2}\Big(W_{\mu\nu\rho\sigma}\pm {}^*W_{\mu\nu\rho\sigma}\Big) ,  
\label{w}
 \end{eqnarray} 
 where
\begin{eqnarray}
{}^*W_{\mu\nu\rho\sigma}=\frac{1}{2} {\epsilon_{\mu\nu}}^{\kappa\lambda}W_{\kappa\lambda\rho\sigma} ~~~~~~\mbox{in}~~~~\MR. 
\end{eqnarray}
In two-component notation we have 
\begin{eqnarray}
  W^+_{\mu\nu\rho\sigma}&=&W_{ABCD}\epsilon_{\dA\dB}\epsilon_{\dot{C}\dot{D}}, 
\nn \\
W^-_{\mu\nu\rho\sigma}&=&W_{\dA\dB\dot{C}\dot{D}}\epsilon_{AB}\epsilon_{CD},
 \label{WPM}
\end{eqnarray}
where $\epsilon_{AB}$ is the $SU(2)$ invariant totally antisymmetric tensor.

In $\ML$ we have a similar   decomposition of the Weyl tensor 
\begin{eqnarray}
W_{\mu\nu\rho\sigma}&=&W^+_{\mu\nu\rho\sigma}+W^-_{\mu\nu\rho\sigma} ,
\nn\\
W^\pm_{\mu\nu\rho\sigma}&=&\frac{1}{2}\Big(W_{\kappa\lambda\rho\sigma}\mp i\, {}^*W_{\mu\nu\rho\sigma}\Big) ,  
\label{w}
 \end{eqnarray} 
where now
\begin{eqnarray}
{}^*W_{\mu\nu\rho\sigma}=\frac{i}{2} {\epsilon_{\mu\nu}}^{\kappa\lambda}W_{\kappa\lambda\rho\sigma} ~~~~~~\mbox{in}~~~~\ML. 
\label{wl}
\end{eqnarray}
In an obvious two-component $SL(2,C)$ notation we have 
\begin{eqnarray}
W^+_{\mu\nu\rho\sigma}&=&
\overline{\Psi}_{\dot{\alpha}\dot{\beta}\dot{\gamma}\dot{\delta}}\epsilon_{\alpha\beta}\epsilon_{\gamma\delta}, \nn \\
W^-_{\mu\nu\rho\sigma}&=& 
\Psi_{\alpha\beta\gamma\delta}\epsilon_{\dot{\alpha}\dot{\beta}}\epsilon_{\dot{\gamma}\dot{\delta}},  \label{WPL}
\end{eqnarray}
where now $(\alpha,\beta,\cdots=1,2)$   are $SL(2,C)$ indices.
 
Let us note that due to the fact that there are no degrees of freedom in the topological phase I, there are no gravitational field equations. Classical configurations are still determined by the critical points of the action (\ref{actionI}). These can be found by recalling that in 
$\MR$, the inequality \cite{SP}
\begin{eqnarray}
\Big(W^{\mu\nu\rho\sigma}\pm {}^*W^{\mu\nu\rho\sigma}\Big)
\Big(W_{\mu\nu\rho\sigma}\pm {}^*W_{\mu\nu\rho\sigma}\Big)\geq 0,
\end{eqnarray}
leads to 
\begin{eqnarray}
{\cal S}_{I}=\frac{1}{g_{\tiny{W}}^2}\int_{{\cal M}_{\rm I}} \d^4 x \sqrt{g}\,   W_{\mu\nu\rho\sigma}
W^{\mu\nu\rho\sigma}\geq \frac{48\pi^2}{g_{\tiny{W}}^2} |\tau_{\tiny{R}}|, \label{actionII}
\end{eqnarray}
where 
\begin{eqnarray}
\tau_{\tiny{R}}=\frac{1}{48\pi^2} \int_{{\cal M}_{\rm I}} \d^4 x \sqrt{g}\,  
 W_{\mu\nu\rho\sigma}
\, {}^*W^{\mu\nu\rho\sigma}  \label{t}
\end{eqnarray}
is the Hirzebruch signature (for a compact manifold). 
Therefore, the topological action is minimized for  self dual configurations 
\begin{eqnarray}
W^-_{\mu\nu\rho\sigma}=W_{\dA\dB\dot{C}\dot{D}}\epsilon_{AB}\epsilon_{CD}=0,
\label{Wp}
\end{eqnarray}
or anti-self dual configurations 
\begin{eqnarray}
W^+_{\mu\nu\rho\sigma}=W_{ABCD}\epsilon_{\dA\dB}\epsilon_{\dot{C}\dot{D}} =0.
\label{Wm}
\end{eqnarray}
 Such configurations are annihilated by the BRST charge since for example  
\begin{eqnarray}
\delta \chi_{ABCD} =i\, W_{ABCD}=0
\end{eqnarray}
on anti-self dual geometry \cite{Vafa}. 
We will now assume that the boundary $B$ is not special in the sense that nothing happens there apart from the change of signature. In particular, there are no discontinuities along the boundary  $B$.  Let us then see what are the consequences  of such an assumption. 
\vskip .1in

    \noindent
 \subsection{Condition on the extrinsic curvature}
Standard Einstein equations  (\ref{EE}) hold in the non-distributional sense in the whole of  $\ML$ including  its boundary. 
That is,  all geometric quantities are continuous at $B$ and in particular,  the second fundamental form (extrinsic curvature) should also be continuous across $B$. Viewed from phase I, i.e., from $\MR$, the second fundamental form is 
\begin{eqnarray}
K_{\mu\nu}=\nabla_\mu n_\nu, 
\label{KR}
\end{eqnarray}
 where $n^\mu$ is the unit normal vector to the boundary $B$. On the other hand,  from the $\ML$ point of view we find 
 \begin{eqnarray}
 K_{\mu\nu}= i\nabla_\mu n_\nu, \label{KL}
 \end{eqnarray}
 since we have to Wick rotate going form $\MR$ to $\ML$. Since the extrinsic curvature is continuous on $B$, the only way both (\ref{KR}) and (\ref{KL}) to hold is 
 \begin{eqnarray}
 K_{\mu\nu}\Big|_B=0.  \label{KB}
 \end{eqnarray}
 In other words, the common boundary of $\MR$ and $\ML$ has vanishing extrinsic curvature and therefore is a totally geodesic submanifold as in the no-boundary proposal \cite{G-H}.\footnote{A hypersurface like $B$ with vanishing extrinsic curvature is also called ``moment of time symmetry'' \cite{Wald}.} This also specifies the scalar curvature of $B$. Indeed, from Einstein equations (\ref{EE}), the Hamiltonian constraint  
 is  
 \begin{eqnarray}
 {}^3R+K^2-K_{ij}K^{ij} =2 M_{\rm p}^{-2} n^\mu n^\nu T_{\mu\nu} ~~~~\mbox{in}~~~\ML
 \end{eqnarray}
and therefore, due to (\ref{KB}), we get that 
\begin{eqnarray}
{}^3R\big|_B=2 M_{\rm p}^{-2} \, \rho \big|_B,  \label{HC}
\end{eqnarray}
where $\rho=n^\mu n^\nu T_{\mu\nu}$ is the energy density.
Since $n^\mu$ is timelike,  the weak energy condition $n^\mu n^\nu T_{\mu\nu}\geq 0$ gives 
\begin{eqnarray}
{}^3R\big|_B\geq 0.  \label{RB}
\end{eqnarray}
Therefore, the induced metric on $B$ has non-negative scalar curvature. 
This result should be kept in mind when discussing the flatness problem.

Since (\ref{HC}) is nothing else than the Hamiltonian constraint, it provides 
initial data for Einstein equations. However, these initial data should not be arbitrary but should describe  a totally geodesic hypersurface  of the  Riemannian space $\MR$, where $\MR$ and $\ML$ are glued in an at least $C^{(2)}$ way.

\noindent
\subsection{Condition on the Weyl tensor}
We have seen that geometries with vanishing self dual (or the anti-self dual) part of the Weyl tensor are critical  points of Witten's topological gravity. Therefore,  at the boundary $B$ of $\MR$ we will have by continuity
\begin{eqnarray}
W_{ABCD}=0 ~~~~~\mbox{in} ~~~~\MR,~~~ \Longrightarrow  ~~~W_{ABCD}\big|_{B=\partial \MR}=0
\end{eqnarray}
as well.  The same condition should hold from the $\ML$ side (phase II) i.e., 
\begin{eqnarray}
W_{ABCD}\big|_{B=\partial \ML}=0.
\end{eqnarray}
Since we have to Wick rotate  from $\MR$ to $\ML$, we have that 
\begin{eqnarray}
W^-_{\mu\nu\rho\sigma}=\frac{1}{2}\Big( W_{\mu\nu\rho\sigma}+{}^*W_{\mu\nu\rho\sigma}\Big)~~~~ \mbox{in} ~~~~\MR,  \nn \\
 W^-_{\mu\nu\rho\sigma}=\frac{1}{2}\Big(W_{\mu\nu\rho\sigma}+i{}^*W_{\mu\nu\rho\sigma}\Big)~~~~ \mbox{in} ~~~~\ML.  \label{Wll}
\end{eqnarray} The only way to match on their common boundary is the full Weyl tensor to vanish, i.e., 
\begin{eqnarray}
W_{\mu\nu\rho\sigma}\big|_B=0.   \label{WW}
\end{eqnarray}
Let us note that the above condition (\ref{WW}) can also be written as
initial data for the electric and magnetic part of the Weyl tensor, which will be useful below, as follows.   
The boundary hypersurface $B$ is spacelike with 
timelike normal $n^\mu$. We can decompose the Weyl tensor into its electric and magnetic parts with respect to $n^\mu$  as 
\begin{eqnarray}
  E_{\mu\nu}&=&W^\rho_{\mu\sigma\nu} n_\rho n^\sigma, \label{El} \\
  B_{\mu\nu}&=&\frac{1}{2} {\epsilon_{\mu\rho}}^{\kappa\sigma}
  W^\lambda_{\nu\kappa\sigma} n^\rho n_\lambda . \label{M}
  \end{eqnarray}  
In terms of the projection $h_{\mu\nu}$ orthogonal to $n^\mu$  
\begin{eqnarray}
h_{\mu\nu}=g_{\mu\nu}+n_\mu n_\nu , \label{h}
\end{eqnarray}
Eqs.(\ref{El}) and (\ref{M}) are  expressed as 
\begin{eqnarray}
E_{\mu\nu}&=&-{}^3R_{\mu\nu}+K K_{\mu\nu}-K_{\mu\lambda}K^\lambda_\nu+
\frac{1}{2} h^\rho_\mu h^\sigma_\nu R_{\rho\sigma}
+\left(\frac{1}{2} h^{\rho\sigma}R_{\rho\sigma}-\frac{1}{3} R\right) h_{\mu\nu}, \label{Ee}\\
B_{\mu\nu}&=& D_\rho K_{\mu(\sigma}{\epsilon_{\nu)}}^{\sigma\rho\kappa}n_\kappa, 
\label{Be}
\end{eqnarray}
where $D_\rho$ is the covariant derivative with respect to the induced metric $h_{\mu\nu}$. 
The vanishing of the Weyl tensor on $B$ then implies 
\begin{eqnarray}
E_{\mu\nu}\big|_B=B_{\mu\nu}\big|_B=0.
\end{eqnarray}
Since the extrinsic curvature of $B$ vanishes according to Eq.(\ref{KB}),
the magnetic part of the Weyl tensor is automatically zero. On the other hand, the vanishing of the electric part  gives that the boundary $B$ is a space of constant curvature \cite{Tod99}. 

There is a second condition related to the Weyl tensor, namely, the Bach tensor should vanish in $\MR$. Indeed, the variation of (\ref{actionI}) leads to 
field equations of conformal gravity 
\begin{eqnarray}
{\cal B}_{\mu\nu}=\left(\nabla^\rho\nabla^\sigma+\frac{1}{2}R^{\rho\sigma}\right) W_{\mu\nu\rho\sigma}=0, \label{Bach}
\end{eqnarray}
where ${\cal B}_{\mu\nu}$ is the Bach tensor.
If the Weyl tensor is self dual or anti-self dual, then the Bach tensor vanishes identically. Indeed, in two-component notation the Bach tensor is written as 
\begin{eqnarray}
{\cal B}_{\mu\nu}&=&2\Big({\nabla^C}_{\dA} {\nabla^D}_{\dB}+{\Phi^{CD}}_{\dA\dB} 
\Big) W_{ABCD}\nn \\
&=&2\Big({\nabla^{\dot{C}}}_{A} {\nabla^{\dot{D}}}_{B}+{\Phi^{\dot{C}\dot{D}}}_{AB} 
\Big) W_{\dA\dB\dot{C}\dot{D}}, \label{BB}
\end{eqnarray}
where 
$\Phi_{AB\dA\dB}=-\frac{1}{2}\big(R_{\mu\nu}- \frac{1}{4}R g_{\mu\nu}\big)$ is the Ricci spinor.  
Clearly ${\cal B}_{\mu\nu}=0$ for self dual ($W_{\dA\dB\dot{C}\dot{D}}=0$), or 
anti-self dual  $(W_{ABCD}=0$)   Weyl. Therefore,
the Bach tensor should vanish in $\MR$ and  by continuity, it 
 should vanish also on the boundary $B$, i.e., 
\begin{eqnarray}
{\cal B}_{\mu\nu}\big|_B=0. \label{BachB}
\end{eqnarray}
The  vanishing of the  Bach tensor on the boundary $B$ should also be true for observers in $\ML$. 

\section{The  cosmological problems in the topological early universe cosmology}

Here we will consider the usual cosmological problems in view of the pre-existence of a topological phase of gravity. In particular, we will examine the horizon and the flatness problem.

\subsection{Horizon Problem}
A serious shortcoming of the standard big bang cosmology  is the horizon problem, which is associated to  the puzzling homogeneity of the observed universe across patches which had never been in causal contact since the onset of the cosmological evolution.  
In the present scenario, all regions of the universe have emerged from a single phase I. As can be seen from Fig. 1, two past light cones, 
although have never been in contact in $\ML$, have emerged, through the boundary $B$,  from the common phase I of the Riemannian $\MR$ space. Therefore, their homogeneity is the result of the same universal initial conditions set from phase I. 

This is also related by the fact that the initial data for the cosmological evolution in phase II in $\ML$, as we have seen, is the vanishing of the 
Weyl tensor and the extrinsic curvature on the common boundary $B$.   

Let us recall that it is generally accepted that an initial vanishing Weyl tensor, a hypothesis that is usually referred to as ``the Weyl curvature hypothesis'', suffices to explain the isotropy and homogeneity  of the universe \cite{PenroseI,PenroseII,PenroseIII}.
In fact, it has been conjectured that an  initially zero Weyl tensor necessary  implies an FRW cosmology.  Although a  general proof is lacking, this conjecture has been proved for a universe filled with a perfect fluid with equation of state $P=w\rho$ for $0< w\leq 1$ \cite{Newman,NewmanII,Anguige,AnguigeII, Tod99}.


\subsection{Flatness}

Although it seems that the horizon problem and the associated homogeneity and isotropy of the universe can easily be explained withing the present set up, this is not automatically the case for the flatness problem. Indeed, as we have seen above, the conditions 
 Eqs. (\ref{KB}) and (\ref{WW})   lead to a constant curvature 3D boundary $B$, see equation (\ref{RB}). 
This poses a threat to the solution to the flatness problem as both
  possibilities $k=0$ and $k=1$ of the spatial geometry of the FRW universe are 
allowed.

One may hope that  a flat $B$ ($k=0$) can be selected by adding an
appropriate  theory on $B$. Such a boundary theory should be conformal as we want the conformal invariance to be broken only by anomalies. Luckily, such a theory exists and  is 
a 3D conformally invariant version of conventional gravity of Chern-Simons type with action 
\cite{DJT1,DJT2,HW}
\begin{eqnarray}
 {\cal S}_{HW}=\int_B \epsilon^{ijk}\left\{\omega_{ia}\big(\partial_j\omega_k^a- \partial_k\omega_j^a\big)+\frac{2}{3} \epsilon^{abc}
 \omega_{ia}\omega_{jb}\omega_{kc}\right\}, \label{HW}
 \end{eqnarray} 
where ${\omega_i}^a$ is the spin connection. As it has been proven in \cite{HW},  3D conformal gravity described by (\ref{HW}) is classically equivalent to a Chern-Simons theory for the conformal group $SO(3,2)$. 
Variation of (\ref{HW}) gives 
\begin{eqnarray}
\delta {\cal S}_{HW}=\int_B \epsilon^{ijk} {R_{ij}}^{ab}\delta \omega_{kab}.
\end{eqnarray}
In the usual treatment of the gravittional Chern-Simons action (\ref{HW}), one trades the variation of the spin connection with that of the vierbeins. This leads to the  
equation of motion 
\begin{eqnarray}
C_{ijk}=\nabla_{k}W_{ij}-\nabla_{j}W_{ki}, 
\end{eqnarray}
where $W_{ij}=R_{ij}-1/4 R g_{ij}$, i.e., to the vanishing of the 3D Cotton tensor
$C_{ijk}$.
Therefore, one is tempting to conclude that $B$ should be only conformally flat  (due to the vanishing of the Cotton tensor), but not necessarily flat ($k=0$). However, in our case, $B$ is the boundary of both $\MR$ and  $\ML$ so that variations of the vierbein vanishes 
\begin{eqnarray}
\delta {e_i}^a\big|_B=0. 
\end{eqnarray}
This means that we should treat the spin connection $\omega_{iab}$ as independent field as in Palatini formulation, so that its equation of motion is 
\begin{eqnarray}
{R_{ij}}^{ab}=0. \label{RR1}
\end{eqnarray}
Therefore we  see that the addition of the conformal invarinat 3D action Eq. (\ref{HW}) selects a flat boundary $B$ solving the flatness problem.

\section{Cosmological Perturbations}

We would like now to characterize the allowed curvature perturbation in $\ML$  given the initial data 
(\ref{WW}) and (\ref{KB}). In other words, we are looking for  perturbations that have vanishing Weyl tensor and extrinsic curvature on $B$ , i.e., perturbations $\delta g_{\mu\nu}$ such that\footnote{The boundary term (\ref{HW}) does not contribute to the  perturbations at the boundary.}
\begin{eqnarray}
\delta E_{ij}\big|_B=0, ~~~~\delta B_{ij}\big|_B=0  \label{per}
\end{eqnarray}
and 
\begin{eqnarray}
\delta K_{ij}\Big|_B=0 . \label{perK}
\end{eqnarray}
A perturbed FRW universe is written in conformal Poisson gauge as 
\begin{eqnarray}
\d s^2=a(\eta)^2 \left[ -(1+2\Phi)\d \eta^2+2 \omega_i \d \eta \d x^i+
\big[(1-2\Psi) \gamma_{ij} +h_{ij}\big] \d x^i \d x^j \right],
\end{eqnarray}
where $\omega_i$ is transverse,  
$h_{ij}$ is transverse and traceless and $\gamma_{ij}$ is the metric with curvature $k=\pm 1,0$. 
Then we find that the associated non-vanishing perturbations of the electric and magnetic 
parts of the Weyl tensor are \cite{Stewart}
\begin{itemize}
\item {\it Scalar perturbations:}
\begin{eqnarray}
\delta E^S_{ij}&=&\frac{1}{2} \left(\nabla_i \nabla_j -\frac{1}{3} \gamma_{ij} \nabla^2\right)\big(\Phi+\Psi\big), \label{sca1} \\
\delta B^S_{ij}&=&0. \label{sca}
\end{eqnarray}
\item  {\it Vector perturbations:}
\begin{eqnarray}
\delta E^V_{ij}&=&\frac{1}{2} \nabla_{(i}\omega'_{j)}, \label{vec1} \\
\delta B^V_{ij}&=&\frac{1}{2}{\epsilon_i}^{mn}\left(
\nabla_j\nabla_m \omega_n-\frac{1}{2}\gamma_{jm} \nabla^2 \omega_n\right). \label{vec}
\end{eqnarray}
\item {\it Tensor perturbations:}
\begin{eqnarray}
\delta E^T_{ij}&=&-\frac{1}{2} h_{ij}''-\frac{1}{2} \nabla^2 h_{ij} +k h_{ij}, ~~~~~~~~~~~\label{ten1} \\
\delta B^T_{ij}&=&-{\epsilon_i}^{mn} \nabla_m h'_{n j}. \label{ten}
\end{eqnarray}
\end{itemize}
It is clear that there are no non-trivial tensor and vector perturbations since the conditions (\ref{per}) specifies the initial second and first derivatives of $h_{ij}$ and $\omega_i$, respectively,  on $B$ 
\begin{eqnarray}
0&=&\left.\left(-\frac{1}{2} h_{ij}''-\frac{1}{2} \nabla^2 h_{ij} +k h_{ij}\right)\right|_B , \label{EBB1} \\
 0&=&\nabla_{(i}\omega '_{j)}\Big|_B,  \label{EBB}
\end{eqnarray}
so that the corresponding Cauchy problem has no solution. 
However,  there are scalar perturbations which should satisfy the initial 
condition
\begin{eqnarray}
\big(\Phi+\Psi\big)\Big|_B=0, \label{pp}
\end{eqnarray}
as can be seen from Eq. (\ref{sca1}).
On the other hand, it  follows from the perturbed Einstein equations that 
\begin{eqnarray}
\Phi-\Psi=-8\pi G \delta \sigma, \label{sig}
\end{eqnarray}
where $\delta \sigma$ is the anisotropic stress. Therefore, from Eqs. (\ref{pp}) and (\ref{sig}) we find that there must exist an initial anisotropic stress given by 
\begin{eqnarray}
\delta \sigma\big|_B=\frac{1}{4\pi G} \Psi\big|_B. 
\end{eqnarray}
However, since shear viscosity is redshifted like $a^{-6}$, an initial $\delta \sigma$ promptly decays. It remains though to  explain its origin from phase I.
One concludes that only scalar perturbations are possible  in the FRW background in  the two-phase gravity. Next, we will determine the power spectum of scalar perturbations.

\subsection{Scalar Perturbations \label{sper}}

We are interested now in scalar perturbations in $\MR$. Such perturbations will provide initial conditions for the two-point correlators at the  Cauchy surface $B$ and determine therefore 
the spectrum of curvature perturbations in the  $\ML$ universe.     
As long as the theory in $\MR$ is conformal,   conformal transformations of the metric 
\begin{eqnarray}
\delta g_{\mu\nu}=2h g_{\mu\nu} 
\end{eqnarray}
do not change the theory.
 Therefore, the conformal
mode corresponding to the conformal perturbation
\begin{eqnarray}
g_{\mu\nu}=\big(1+2h\big)\bar{g}_{\mu\nu},~~~~\tau=\bar{\tau}+h, \label{ct}
\end{eqnarray}
 where $\bar{g}_{\mu\nu}$ is the background metric, is not physical.
However, when conformal invariance is broken, the conformal mode becomes physical (dilaton).  In particular, if conformal invariance is broken due to an anomaly, the trace of the energy-momentum tensor is given by

\begin{eqnarray}
T^\mu_\mu=\frac{c}{1920\pi^2}W_{\mu\nu\rho\sigma}W^{\mu\nu\rho\sigma}-
\frac{a}{5760\pi^2} E_4,
\end{eqnarray}
normalized such that  a real  scalar contribute with $a=c=1$ to the anomaly.
  The response of the effective action under (\ref{ct}) is 
\begin{eqnarray}
\delta_{h} {\cal S}=\int \d^4 x \sqrt{g} \, h\left(\frac{c}{1920 \pi^2} W_{\mu\nu\rho\sigma}W^{\mu\nu\rho\sigma}-\frac{a}{5760\pi^2} E_4\right).
\end{eqnarray}
Then, it is straightforward to verify that  
\begin{eqnarray}
{\cal S}_{ h}=-\frac{a}{720 \pi^2}\int \d^4 x \sqrt{\bar{g}}\bigg\{ \bar{G}^{\mu\nu}\, \partial_\mu h\partial _\nu h+\bar{R}^{\mu\nu}
h\partial_\mu h\partial _\nu h +\partial_\mu h\partial_\nu h\partial^\mu h\partial^\nu h +\mathcal{O}(h^5)
\bigg\}, \label{Ga}
\end{eqnarray}
to fourth order where $\bar{G}^{\mu\nu}$ ($\bar R ^{\mu\nu}$) the background Einstein (Ricci)  tensor. Therefore, the conformal mode is dynamical now. Its two-point correlator can be found by using the Ward identity for scale invariance. The latter is written as 
\begin{eqnarray}
\langle T^\mu_{\mu}(x) {\cal O}_1(x_1)\cdots {\cal O}_n(x_n)\rangle
=-\sum_{i=1}^n \delta^{(4)}(x-x_i) \Delta_i \langle  {\cal O}_1(x_1)\cdots
{\cal O}_i(x_i)\cdots  {\cal O}_n(x_n)\rangle, \label{Ward}
\end{eqnarray}
where $\Delta_i$ is the dimension of ${\cal O}_i$. In particular, for ${\cal O}_i=h$, the Ward identity (\ref{Ward}) for  the two-point correlator 
is written as 
\begin{eqnarray}
\langle T^\mu_{\mu}(x) h(y)h(0)\rangle
=- \Delta\, \delta^{(4)}(x-y) \langle   h(y)h(0)\rangle - 
\Delta\, \delta^{(4)}(x) \langle   h(y)h(0)\rangle. 
\end{eqnarray}
Therefore, after integration we get that 
\begin{align}
2\Delta \langle   h(y)h(0)\rangle &=-\int_{\MR} \d^4 x \sqrt{g} \langle T^\mu_{\mu}(x) h(y)h(0)\rangle\nonumber\\
&\approx -\int_{\MR} \d^4 x\sqrt{g}\,
\langle T_{\mu\nu}(x)\rangle 
 \langle   h(y)h(0)\rangle . \label{tmm}
\end{align}
where the approximation $\langle T^\mu_{\mu}(x) h(y)h(0)\rangle
\approx \langle T^\mu_{\mu}(x)\rangle\langle h(y)h(0)\rangle$
has been used.
%
%
Using the integrated conformal anomaly 
\begin{eqnarray}
\int_{\MR} \d^4 x \sqrt{g} \big<T^\mu_\mu(x)\big>=\frac{c}{240} 
\langle I_W\rangle  -
\frac{a}{180} \chi_{\tiny{R}}, 
\end{eqnarray}
where $ \chi_{\tiny{R}}$ is given by 
\begin{eqnarray}
\chi_{\tiny{R}}=\frac{1}{32\pi^2} \int\limits_{\MR} E_4,  \label{euler}
\end{eqnarray}
and  $I_W$ denotes the integral
\begin{eqnarray}
I_W=\frac{1}{8\pi^2}\int\limits_{\MR} W_{\mu\nu\rho\sigma}^2\, \sqrt{g} 
\d^4 x,
\end{eqnarray}
we find that  the scaling dimension of $h$  turns out to be
\begin{eqnarray}
\Delta\approx \frac{a}{360} \chi_{\tiny{R}}-\frac{c}{480} 
\langle I_W\rangle.
\end{eqnarray}
 Note that $\chi_{\tiny{R}}$ is the Euler number for a compact 4D manifolds. 
If there are boundaries, then the Euler number gets boundary corrections 
\cite{EGH} (and so does the integrated conformal anomaly
\cite{Dowker, Solo}) so that 
\begin{eqnarray}
\chi_{\tiny{R}}=\frac{1}{32\pi^2} \int\limits_{\MR} E_4-\frac{1}{4\pi^2}
\int\limits_{\partial \MR} X,
\end{eqnarray}
where $X$ is 
\begin{align}
X&=K^{\mu\nu}n^\kappa n^\lambda R_{\kappa\mu\lambda\nu}-K^{\mu\nu}R_{\mu\nu}
-K R_{\mu\nu}n^\mu n^\nu \nonumber \\
&+\frac{1}{2}K R-\frac{1}{3}K^3
+K K^{\mu\nu}K_{\mu\nu}+\frac{2}{3}
K^{\mu\kappa}K_{\kappa\lambda}K^\lambda_\nu. 
\end{align}
In the present case, since the extrinsic curvature $K_{\mu\nu}$ vanishes, we have $X=0$ and therefore, there are no boundary corrections to the Euler number. In other words, the Euler number of $\MR$ with boundary the totally geodesic codimension-one space $B$, is given just by  (\ref{euler}).

An estimate of the term $\langle I_W\rangle$ gives  
\begin{eqnarray}
\langle I_W\rangle \sim \alpha_W 
\Lambda_{UV}^4 L_I^4, \label{vv}
\end{eqnarray}
where $\Lambda_{UV}$ is  the UV cutoff in $\MR$, 
$L_I$ is a characteristic scale in $\MR$ (curvature scale) 
and $a_W=g_W^2/8\pi$.  
 Indeed,  based on dimensional grounds,  $\langle W_{\mu\nu\rho\sigma}^2 \rangle\sim \alpha_W \Lambda_{UV}^4$ and 
  therefore, 
we expect   (\ref{vv}) to hold. 


Note that we  may use the doubling trick to construct a new compact manifold out of $\MR$ if its boundary is only $B$. The rule is the following: 
since the extrinsic curvature of $\MR$ vanishes, we may 
 construct the manifold $2\MR$ consisting of two copies $\MR^-$ and $\MR^+$ 
 joined across $B$. Since $K_{ij}=0$ on $B$, the induced metric on $2\MR$ from 
 the metrics on $\MR^\pm$ is at least $C^{(1)}$. The manifold $2\MR$ is a compact manifold without boundary, it has $\chi_R(2\MR)=2\chi_R(\MR)$ and  signature $\tau_R(2\MR)=0$ \cite{G-H}. 
 In addition, $\MR$ admits a reflection map $\theta$ interchanging $\MR^+$ and $\MR^-$ while leaving $B$ invariant. In this case, the metric close to the boundary $B$ is of the form 
 \begin{eqnarray}
 \d s_{\rm I}^2\approx \d t_{\rm I}^2+g_{ij}(x,t_{\rm I}^2)\d x^i\d x^j ,
 \end{eqnarray}
where the reflection map is realized by $t_{\rm I}\to -t_{\rm I}$. Note also that this  form of the metric close to $B$ allows the Wick rotation $t_{\rm I}\to i t_{\rm II }$.
Then the boundary $B$ is just the fixed point of the reflection map 
$\theta$ (acting here as $t_{\rm I} \to -t_{\rm I}$). 

The two-point function turns out to be 
\begin{eqnarray}
\langle   h(x)h(0)\rangle\sim  \frac{720\pi^2L_I^2}{a} \frac{1}{|x|^{2\Delta}},
\end{eqnarray}
and therefore,  we have at the boundary $B$
\begin{eqnarray}
\langle   h(x)h(0)\rangle\Big|_B=\langle   h(0,\vec{x})h(0,\vec{0})\rangle\sim  \frac{720\pi^2 L_I^2}{a} \frac{1}{|\vec{x}|^{2\Delta}}. \label{aa}
\end{eqnarray}
This is  the initial condition for the scalar perturbation in $\ML$. 
The factor $L_I^2/a$ originates from the coupling of the conformal mode in (\ref{Ga}). 
In particular,  $h|_B=\Psi|_B$  and, after Fourier transforming, we 
get that 
the initial condition for the two-point correlator in $\ML$ should be 
\begin{eqnarray}
|\Psi_k|^2\Big|_B\sim \frac{L_I^2 M_{\rm p}^2}{a}  k^{-3+2\Delta}, 
\label{spe}
\end{eqnarray}
since $\Psi$ and $h$ are differently normalized as their kinetic terms are multiplied with $M_{\rm p}^2$ and $L_I^{-2}$, respectively. 
This leads to  a spectral index
\begin{eqnarray}
n_s=1+2\Delta=1+\frac{a}{180} \chi_{\tiny{R}} -\frac{c}{240 } \,\alpha_W
\Lambda_{UV}^4 L_I^4 
\label{ns}
\end{eqnarray}
and an amplitude 
\begin{eqnarray}
A\sim \frac{L_I^2 M_{\rm p}^2}{a}.  \label{A}
\end{eqnarray}
We encounter now a problem related to the  spectral index 
in Eq. (\ref{ns}). Consider the simplest most symmetric case of an $S^4$ of radius $L$ as a candidate for $2\MR$
and  the standard metric on $S^4$
\begin{eqnarray}
\d s^2=L^2\d \tau^2+L^2 \cos^2\tau\, \d\Omega_3, ~~~~-\frac{\pi }{2}\leq \tau
\leq \frac{\pi }{2},
\end{eqnarray}
where $\d \Omega_3$ is the metric of the unit three-sphere and we have shifted the azimuthial angle $\tau$ by $-\pi/2$. Then, the equatorial $\tau=0$ has vanishing extrinsic curvature $K_{ij}=0$ and therefore, it is the boundary $B$ which will be continued into a Lorentzian 
FRW $\ML$ with $k=1$. However, we  find that 
in this case, $\chi_{\tiny{R}}=1$, and  therefore we will have a ridiculously large blue spectrum as follows from (\ref{ns}) and (\ref{A}). 
Now,  in order to get an amplitude  $A\sim 10^{-10}$, we should have $a\sim  10^{10} L_I^2 M_{\rm p}^2$. In addition, we need 
\begin{eqnarray}
-\frac{a}{180} \chi_R\approx 0.03 \label{tx}
\end{eqnarray}
for a red spectrum with $n_s\approx 0.97$. However, according to Hopf conjecture, the Euler number of a compact manifold is no-negative, so that
\begin{eqnarray}
\chi_R(\MR)=\frac{1}{2}\chi_R(2\MR)=1-b_1+\frac{1}{2}b_2\geq 0
\end{eqnarray}
where $b_1,b_2$ are the Betti numbers of $2\MR$ \cite{G-H}. Therefore, we end up in general in an enormous blue spectrum for scalar perturbations. 
  The only way  to avoid this is the manifolds $\MR$ to have vanishing 
  Euler number  
\begin{eqnarray}
\chi_R=0. \label{xt}
\end{eqnarray}
 In this case, the spectral index is given by 
\begin{eqnarray}
n_s=1+2\Delta=1 -\frac{c}{480} \,\alpha_W
\Lambda_{UV}^4 L_I^4, 
\label{ns1}
\end{eqnarray}
which may have  the correct value for appropriate values of the parameters \cite{Vafa}.

Compact manifolds with $\chi_R=0$ admit codimension-one foliations \cite{thurston}. In addition, the leaves of the foliation are totally geodesic 
(vanishing extrinsic curvature) if $\tau_R=0$ as well \cite{JN}. 
In other words, compact manifolds satisfying (\ref{xt}) can be cut along appropriate leaf  
$B$ and continued to $\ML$ with Lorentzian signature. However, $\MR$ should also be self-dual. Examples of such spaces include $T^4$ and $S^1\otimes S^3$ and connected sums of these which are (trivially) self-dual
and satisfy (\ref{xt}). The boundary in this case can be either $T^3$ or $S^1\otimes S^2$. The latter case however, is excluded as it leads to  a not flat boundary, as required
by Eq.(\ref{RR1}). 

Non-compact manifolds with $\chi_R=0$ can  be constructed. 
For example, let us assume that the metric of $2\MR$ is of the form 
\begin{eqnarray}
\d s^2=\d \tau^2+a^2(\tau) \delta_{ij}(x)\d x^i \d x^j,  ~~~~-\tau_0\leq \tau\leq \tau_0,
\label{ex}
\end{eqnarray}
where 
$\tau_0$ is a real constant. It is conformally flat, and therefore trivially self-dual (vanishing Weyl tensor). 
The boundary $B$ is the hypersurface
at  $\tau=0$. Its extrinsic curvature is $$K_{ij}=a'(0)\delta_{ij}$$ and therefore, 
$B$ is totally geodesic if $a'(0)=0$. Its Euler number 
$\chi_R$ turns out to be 
\begin{eqnarray}
\chi_{\tiny{R}}=\frac{V}{4\pi^2}a'(\tau)^3,
\end{eqnarray}
Then,  $\chi_R=0$ is satisfied for 
$a'(-\tau_0)=0$. 


One can also  estimate the  non-linear parameters of  the three- and four-point functions of the comoving curvature perturbation $\zeta=5\Psi/3$ \cite{Vafa}. In the squeezed limit of the three-point function $(k_1\ll k_2\sim k_3$), we have (the prime indicates we do not write the Dirac delta for the momentum conservation and the $(2\pi)^3$ factors) \cite{ngreview}
\begin{eqnarray}
 \Big< \zeta_{\vec{k}_1}\zeta_{\vec{k}_2}\zeta_{\vec{k}_3}\Big>'
 =\frac{12}{5} f_{NL}P^\zeta_{\vec{k}_1}P^\zeta_{\vec{k}_2},
 \end{eqnarray} 
 where $P^\zeta_{\vec{k}}$ is the power spectrum deduced by  Eq. (\ref{spe}). On the other hand, for the four-point function in the collapsed limit ($\vec{k}_{12}=\vec{k}_{1}+\vec{k}_{2}\approx 0$), we have 
 \begin{eqnarray}
 \langle \zeta_{\vec{k}_1}\zeta_{\vec{k}_2}\zeta_{\vec{k}_3}\zeta_{\vec{k}_4}\rangle'
 =4 \tau_{NL} P^\zeta_{\vec{k}_1}P^\zeta_{\vec{k}_3}P^\zeta_{\vec{k}_{12}}, 
 \end{eqnarray}
 whereas, in the squeezed limit $(k_4\ll k_1,k_2,k_3$),  
 \begin{eqnarray}
 \Big< \zeta_{\vec{k}_1}\zeta_{\vec{k}_2}\zeta_{\vec{k}_3}\zeta_{\vec{k}_4}\Big>'
 =\left(2 \tau_{NL} +\frac{54}{25}g_{NL}\right)P^\zeta_{\vec{k}_4}\Big(P^\zeta_{\vec{k}_1}P^\zeta_{\vec{k}_{12}}+2\, \mbox{perm.}\Big).
 \end{eqnarray}
 From the action (\ref{Ga}), it can easily be verified that the three- and 
 four-point functions scale as 
 \begin{eqnarray}
 \Big< \zeta_{\vec{k}_1}\zeta_{\vec{k}_2}\zeta_{\vec{k}_3}\Big>'\sim \frac{L_I^4 M_{\rm p}^4}{a^2}, ~~~~
 \Big<\zeta_{\vec{k}_1}\zeta_{\vec{k}_2}\zeta_{\vec{k}_3}\zeta_{\vec{k}_4}\Big>'\sim  \frac{L_I^8 M_{\rm p}^8}{a^3}. 
 \end{eqnarray}
By using that $P^\zeta_{\vec{k}}\sim L_I^2 M_{\rm p}^2/a$, we find 
\begin{eqnarray}
\label{ng}
f_{NL}\sim {\cal O}(1), ~~~~~\tau_{NL}\sim g_{NL}\sim L_I^2 M_{\rm p}^2. 
\end{eqnarray}
Therefore, since the IR scale $L_I>M_{\rm p}^{-1}$, the Suyama-Yamaguchi inequality \cite{SY1,SY2}
\begin{eqnarray}
\tau_{NL}\geq \left(\frac{6}{5}f_{NL}\right)^2, 
\end{eqnarray}
is satisfied for this class of models.

To summarise, we have found that the Riemannian manifold $\MR$
satisfy the following conditions: 
\begin{itemize}
\item It has self-dual Weyl tensor (vacuum of the topological theory on $\MR$.)
\item It accommodates a codimension-one totally geodesic submanifold $B$ (zero  extrinsic curvature).
\item It  admits  a reflection map (at least locally close to $B$) such that the Wick rotation to the Lorentzian $\ML$ is possible.
\item It has  vanishing Euler number ($\chi_R=0$) in order to have a red 
spectrum of scalar perturbations. 
\end{itemize} 
Importantly, we have seen that the space of manifolds $\MR$ satisfying the above criteria is not empty.  

\vskip 1cm 

\section{Relation with other proposals}

The two-phase gravity is related also to other proposals for the early history of the universe. In particular, it is related to the Weyl curvature hypothesis, the no-boundary program and the out-of nothing creation of the universe. We will examine here the relation of the two-phase model to the aforementioned proposals.

\subsection{Weyl curvature hypothesis}

The Weyl curvature hypothesis has been proposed by R. Penrose motivated by the second law of thermodynamics \cite{PenroseI,PenroseII,PenroseIII}. According to it,  since the entropy of the universe increases with time 
it had a tiny value in the past.
 Penrose noticed  an apparent paradox related with this. 
The best evidence for  Big Bang arises from observations of the CMB. The latter follows  Planck's law to  extraordinary precision, leading to the 
unavoidable conclusion that the early universe was in   thermal equilibrium. But thermal equilibrium represents maximum randomness and therefore it  corresponds to maximum entropy.  How such a picture could be correct given the fact that the universe has started in a low entropy state?
Penrose answers this puzzle by noticing that the   high entropy of the CMB is related to 
the matter content of the universe only and not to gravity. 
In other words, the extraordinary uniformity of the early universe is attributed just to matter while  gravitational degrees of freedom, although potentially available, have not been excited.  The moment gravitational degrees of freedom are activated, the entropy starts to increase due to the clumping of the  initially distribution of matter. 

But how gravitational degree of freedom can be inactivated? According to Penrose, this may be implemented by assuming a principle which he named ``the Weyl curvature hypothesis". Let us recall that curvature is not completely specified in Einstein gravity. Indeed, matter distribution  can only determine the  Ricci curvature once the energy-momentum tensor is known. The rest piece of 
Riemann curvature  is just the Weyl tensor ${W^\mu}_{\nu\rho\sigma}$  which is not specified by Einstein equations. Weyl tensor describes pure gravitational dof and the Weyl curvature hypothesis asserts that this tensor vanishes at the initial Bing-Bang singularity, i.e., 
\begin{eqnarray}
{W^\mu}_{\nu\rho\sigma}\big|_{\tiny{\rm{Big-Bang}}}=0.
\end{eqnarray}
Clearly, this is just our  (\ref{WW}), when the boundary $B$ is identified with the moment of Big Bang. This of course should be expected as phase I has no gravitational degrees of freedom and therefore, necessarily, the Weyl tensor should vanish on any hypersurface that separates phase I with a second phase where Einstein equations hold. In other words, the fact that phase I is topological means that when continued to phase II, the Weyl curvature hypothesis should valid. However, in the case of Penrose, the hypersurface 
$B$ is just a part of an infinite sequence of universes, the ``aenos", making up the Cyclic Conformal Cosmology \cite{PenroseIV},  according to which the universe undergoes repeated  expansion cycles, named aeons, such that each one starting from its own Big-Bang, ending in a  a stage of accelerated expansion and  continues indefinitely. 

In the two-phase proposal on the other side, there are no infinitely many past universes. The latter are  replaced with  a single topological phase
where geometry is Riemannian and there are no  gravitational degrees of freedom. In this respect it seems similar  to the out of nothing creation of the universe which we will now turn.

\subsection{ Creatio Ex Nihilo}

Cosmological spacetimes, even inflationary ones,  are past-incomplete under general assumptions. Therefore, such spacetimes should have a past boundary where initial conditions should be defined. Clearly, a question that should be asked in this case, is what determines these initial conditions. Here, 
following \cite{Vafa}
we have
consider the case in which the universe had two phases: a topological phase I where gravity is not dynamical and the geometry is Riemannian and a dynamical phase Einstein II, where gravity is propagating and the geometry is pseudo-Riemannian specified by Einstein equations. 
However,  motivated by quantum cosmology, it may be that the universe  can spontaneously nucleated out of nothing \cite{VilenkinI,VilenkinII}. This is an old idea and it is based on non-perturbative tunneling where one is looking 
for a ``bounce'' solution of the classical euclidean equations. This is a solution which  approaches the putative vacuum state asymptotically, at infinity. A particular bounce is the de Sitter instanton \cite{G-H}, which has metric 
\begin{eqnarray}
\d s^2=\d \tau^2 +\frac{1}{H^2}\cos^2\big(H\tau\big)\left(
\frac{\d r^2}{1-r^2}+r^2 \d \Omega_2^2\right),  \label{vil}
\end{eqnarray}
and describes the round four-sphere $S^4$. At  $\tau=0$  we have a plane of symmetry and we can rotate $\tau\to it$ where the geometry turns out to be that of standard de  Sitter  with metric 
\begin{eqnarray}
\d s^2=-\d t^2 +\frac{1}{H^2}\cosh^2\big(Ht\big)\left(
\frac{\d r^2}{1-r^2}+r^2 \d \Omega_2^2\right).  \label{vil2}
\end{eqnarray}
The picture looks similar with that of fig 1, where $\MR$ is $S^4$ and  $\ML$ is de Sitter joined together at the maximal (totally geodesic) $B=S^3$  at $\tau=0$. However, the interpretation is different. Although (\ref{vil}) indeed  bounces at the classical turning point, there is no any putative vacuum state to approach asymptotically at infinity since $S^4$ is a compact space and $\tau$ is bounded in the range $-\pi/H<\tau<\pi/H$. From this perspective, the de Sitter spacetime   (\ref{vil2}) appears out of nothing,
 since there is no  classical space, time or matter from which de Sitter emerges. However, in the two phase gravity model we advocate here,
$S^4$ describes phase I and de Sitter phase II. So, the out of nothing creation of the universe is not a bounce in the two-phase model but rather a
Riemannian manifold (here $S^4$) sewing with a speudo-Riemannian one (here de Sitter) along a hypersurface of vanishing extrinsic curvature (here $S^3$).
In this respect, the model we are advocating here
fits better to the no-boundary proposal which we now briefly present below.   

\subsection{No-boundary proposal}

The  Hartle-Hawking no-boundary proposal \cite{Hartle-Hawking}  provides  initial conditions of the universe in the sense that it assigns weighted  probabilities among all possible  cosmological evolutions of our universe. Similarly with the Weyl curvature hypothesis and the out of nothing creation of the universe, the Big Bang singularity is replaced with a smoothed geometry. The no-boundary geometry is constructed by gluing two sectors described by a Riemannian manifold with Euclidean signature and a pseudo-Riemannian one with Loretzian signature 
which are solutions of the Euclidean and Lorentzian Einstein field equation, respectively. Then in the presence of other matter fields like scalars, there are complex solutions that interpolate between the two sectors. 
Among these possible complex solutions, particularly important are solutions describing real tunneling metrics. The latter  describes transitions from a purely Euclidean sector to a  Lorentzian one. Semiclassically, such transitions are described 
by a Euclidean instanton glued to  a Lorentzian solution representing a bubble of true vacuum expanding at the speed of light. 
Such gravitational  instantons have the form of Fig.1 and provide initial conditions for its   Lorentzian counterpart, similarly to the present two-phase gravity model we discuss here. In addition, these instantons exhibit  discontinued metrics since the metric changes signature in the two regions, and therefore,   the Lorentzian sector of the tunneling solutions provides background spacetimes for description of the dynamics of the late universe.  

\noindent
	\begin{center}
		{\bf  Acknowledgments}
	\end{center}
	\noindent
	We would like to thank C. Vafa for correspondance and J. Sonner for discussions.
	A.R. is supported by the Swiss National Science Foundation 
	(SNSF), project {\sl The Non-Gaussian Universe and Cosmological Symmetries}, project number: 200020-178787.


\end{document}